PAPER • OPEN ACCESS

# Analysis of genetic diversity and genome relationships of four eggplant species (*Solanum melongena* L) using RAPD markers

To cite this article: Susilo and M Setyaningsih 2018 *J. Phys.: Conf. Ser.* **948** 012017

View the article online for updates and enhancements.

# Analysis of genetic diversity and genome relationships of four eggplant species (*Solanum melongena* L) using RAPD markers

**Susilo**[1] **and M Setyaningsih**[1]
[1]Departement of Biology Education, Universitas Muhammadiyah Prof. DR. HAMKA. Jalan Tanah Merdeka, Kampung Rambutan, Ciracas, East Jakarta, Indonesia, 13830

E-mail: susilo@uhamka.ac.id

**Abstract.** *Solanum melongena* (eggplant) is one of the diversity of the Solanum family which is grown and widely spread in Indonesia and widely used by the community. This research explored the genetic diversity of four local Indonesian eggplant species namely leuca, tekokak, gelatik and kopek by using RAPD (*Random Amplified Polymorphic DNA*). The samples were obtained from Agricultural Technology Assessment Institute (BPTP) Bogor, Indonesia. The result of data observation was in the form of *Solanum melongena* plant's DNA profile analyzed descriptively and quantitatively. 30 DNA bands (28 polymorphic and 2 monomorphic) were successfully scored by using four primers (OPF-01, OPF-02, OPF-03, and OPF-04). The Primers were used able to amplify all of the four eggplant samples. The result of PCR-RAPD visualization produces bands of 300-1500 bp. The result of cluster analysis showed the existence of three clusters (A, B, and C). Cluster A (coefficient of equal to 49%) consisted of a gelatik, cluster B (coefficient of 65% equilibrium) consisted of TPU (Kopek) and TK (Tekokak), and cluster C (55% equilibrium coefficient) consisted of LC (Leunca). These results indicated that the closest proximity is found in samples of TK (Tekokak) and TPU (Kopek).

## 1. Introduction
One of the national problems faced and must be solved the way out was energy and food tenacity. Therefore, global and fundamental thinking were needed in improving agricultural productivity in order to boost food tenacity in Indonesia. One such effort was to increase the production of vegetable or olericulture crops. Eggplant (*Solanum melongena*) is one type of vegetable that needs to be improved because it has many nutrients including protein, carbohydrate, fiber, iron, vitamins and calcium [22, 12]. These plant also contains phytonutrient anthocyanin or nasunin on purple eggplant skin that highly efficacious for nutrition one of the most important organs of our body like the brain [1, 24]. In addition to nourishing the brain, nasunin is also useful for protecting brain cell membranes from fatal harmful [2, 14, 3].

In Indonesia, many eggplant plants are used by the community as vegetable crops such as kopek, gelatik, tekokak, and leunca. These plants do have morphological character differences, but for molecular data between cultivars have not been studied. The data is very important because it is one of the genetic resources of Indonesia which can be used to improve the germplasm of the nation [19, 23, 9]

Given the countless benefits and the abundant number of eggplant cultivars, it appears to be a must to review and catalog the plant's genetic resources [8]. Molecular data collection of Solanum crops is of paramount importance to support further research [16]. It can also be used as a nation-wide database of germplasm collections in Indonesia as well as further studying the benefits of each plant [2]. The large diversity of Solanum crops in Indonesia requires further research to examine its benefits. Based on the above description, the researcher is interested to examine the DNA character of some *Solanum melongena* or eggplants by using RAPD markers

Sequencing-based on random amplified polymorphic DNA (RAPD) provides better resolution at intra-genus and levels above. RAPD can provide a tool for classifying individuals into the nominal genotypic category and are largely suitable for intra-species genotypic variation study [7, 5]. RAPD is based on the amplification of genomic DNA with single primers of arbitrary nucleotide sequence [10, 13]. These primers detect polymorphisms in the absence of specific nucleotide sequence information and the polymorphisms function as genetic markers and can be used to construct genetic maps [14]. The objective of this paper was to use clearly amplified RAPD fragments to examine the genetic variation within and between populations of four local *Solanum melongena* in Indonesia.

## 2. Methods
### 2.1. Plant Collection
A total of fours *Solanum melongena* namely leuca (LC), tekokak (TK), gelatik (TBU) and kopek (TPU) were included in the present investigation. Young leaf tissue was taken on each species and directly inserted into the plastic and stored into the coolbox for further DNA testing in the laboratory [6]. DNA test by RAPD technique was done at Balai Besar and Development of Biotechnology and Genetic Resources of Bogor Agriculture, Indonesia.

### 2.2. DNA Extraction
The eggplant leaf tissue was grinded to powder in liquid nitrogen using a mortar and pestle and then inserted into 2.0 ml microtube. 200 μl CTAB buffer was added that has been heated at 650C for 5 minutes. The sample was incubated in a water bath with a temperature 650C for 1 hour, flipped back and forth every 10 minutes after cooling at normal room's temperature for approximately 10 minutes. The next step was added 60 μl of NaOac, 700 μl Chisam (mixed chloroform: isoamic alcohol with ratio 24:1). Then centrifuged for 10 minutes with a speed of 12,000 rpm at 200C. The supernate obtained was taken 400 μl and transferred to 1.5 ml microtube. 900 μl of Chisam was added and centrifuged for 10 minutes with a speed of 12,000 rpm at 200C. Then Separated 400 μl supernatant and added 1000 μl of absolute ethanol and then centrifuged again for 5 min with speed 12.000 rpm at 200C. The pellet was washed with 70% ethanol to get rid of salts, air dried and dissolved in sterile water and then settled for over a night [20].

### 2.3. The Quality and Quantity of DNA
Quality testing and concentration on DNA isolation are important for the manipulation of all substances. The DNA concentration and it is purity can be determined by spectrometers using UV light at wavelengths of 260 and 280 nm. The comparison in the test using UV light with a length of 260 / 280nm measures a proportion of proteins contamination in a DNA, and the use of UV with a length of 260 nm aims to calculate the concentration of a DNA in the sample. However, the UV absorption process cannot separate RNA contamination and contaminants such as proteins, phenols, oligo or polysaccharides contained in DNA.

### 2.4. RAPD Process
Four arbitrary oligonucleotide primers obtained from Genetic Science Inc. (Singapore). The primers used in this study were OPF-01 (ATGTCACCAC), OPF-02 (TCAGGACTCC), OPF-03 (TCGAGGACCT), OPF-04 (TAGCTTCACG). Pure genomic DNA was amplified using a single primer dekanucleotide. PCR cocktail was used 1 μl DNA template, 2 μl of ultrapure water, 5 μl mix (kappa 2g fast reading mix) and 1 μl primer. PCR was performed under conditions of pre-denaturation at 960C, denaturation at 940C, annealing at 370C, elongation at 720C and termination at 720C. Then the dye loading solution was added to increase the weight of the DNA molecule. The amplification results were visualized using horizontal electrophoresis with 1% agarose gel (w/v) in a 1x TAE buffer. The agarose gel was immersed in EtBr solution so the ribbon pattern could be seen under the ultraviolet Geldoc.

*2.5. Data Analysis*
The results of DNA bands emerging from the RAPD process showed the molecular weight of DNA. The banding pattern profile from DNA analysis was detected using 254 nm UV rays and sprayed with general spray reagent (Cerium (IV) Sulfate). The parameters observed were the presence or absence of the band on the agarose gel after UV irradiation. The emulated RAPD ribbon is assigned a value of "1" and that does not appear given a value of "0". Only clear, consistent bands and polymorphic bands were used to create binary matrices for statistical analysis. The amplification products obtained for all samples were compared to each other. In order to assess the genetic structure within and among eggplant, all 4 diagnostic RAPD markers were analyzed using NTSYS-pc. Genetic similarities were calculated between populations for producing a UPGMA cluster (Unweighted Pair Group Method Averaging).

**3. Results and Discussion**
*3.1. DNA Profile*
DNA quality and quantity test has been done with Spectrophotometry and Agarose Gel Electrophoresis. The amplification results of the total quality and quantity test of eggplant genome with agarose and spectrophotometry resulted in the purity and concentration of DNA seen in Table 1.

**Table 1.** Total number of purity and genomic DNA concentrations

| Sample code | Purity(µg/µl) | Concentrations (nm) |
|---|---|---|
| Gelatik (TBU) | 1,1272 | 1780,24 |
| Kopek (TPU) | 1,2219 | 1196,19 |
| Tekokak (TK) | 1,2316 | 1830,97 |
| Leunca (LC) | 1,5127 | 1266,21 |

Based on Table 1, the result of quantity test using spectrophotometer showed the highest purity result was in LC sample with purity of 1, 5127 M and concentration 1266, 21 nm, while the lowest purity was in TBU sample with purity 1, 1272 M and concentration 1780, 24 nm. Four primers which showed intra-and interspecific banding pattern for these species were used for further analysis of four *Solanum melongena* sample. All samples yield a readable and scored banding pattern, so the results can be used for analysis. All the primers showed interspecific RAPD polymorphism while some primers showed intraspecific variation (Table 2).

**Table 2.** The number of primers used for DNA amplification of genomic DNA and level of polymorphism

| Primers | Sequence | Polymorphic | Monomorphic | Total bands | Polymorphic Percentage |
|---|---|---|---|---|---|
| OPF-01 | ATGTCACCAC | 5 | 0 | 5 | 100 |
| OPF-02 | TCAGGACTCC | 2 | 1 | 3 | 150 |
| OPF-03 | TCGAGGACCT | 11 | 1 | 12 | 100 |
| OPF-04 | TAGCTTCACG | 10 | 0 | 10 | 100 |
|  | Total | 28 | 2 | 30 |  |

*3.2. Species Relationships*

PCR-RAPD amplification of total genomic DNA from four accessions of *Solanum melongena* by using four RAPD primers (OPF-04, OPF-03, OPF-02, OPF-02, and OPF-01) produced 30 DNA bands that could be scored. The number of bands produced ranges from 3 (OPF-02) to 12 (OPF-03) with the resulting band sizes ranging from 300 bp to 2000 bp.

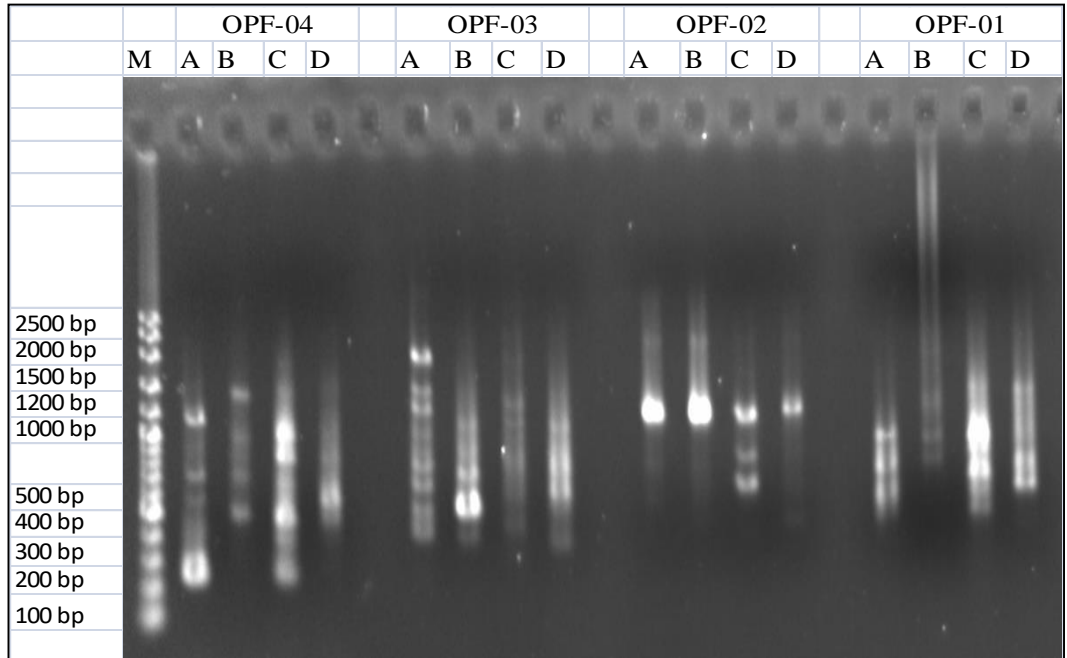

**Figure 1.** RAPD profiles of total genomic DNA from four accessions of *Solanum melongena* amplified by using random primer OPF-04, OPF-03, OPF-02, OPF-01. (M: Marka /Ledder, A: Gelatik, B: Kopek, C: Tekokak and D: Leunca).

Cluster analysis showed the separation of eggplant into clusters clustered by type Figure 3. The coefficient of the genetic similarity of the four accessions of eggplant ranged from 25% to 100%. The result of cluster analysis showed the existence of three clusters (A, B, and C). Cluster A (coefficient of equality of 49%) consisted of a gelatik. Cluster B (coefficient of 65% equilibrium) was consisting of kopek and tekokak. Cluster C (55% equation coefficient) consisted of Leunca (Figure 2). The dendrogram was constructed based on the percentage similarity calculated after pooling the data from 30 loci RAPD of four accessions of *Solanum melongena*.

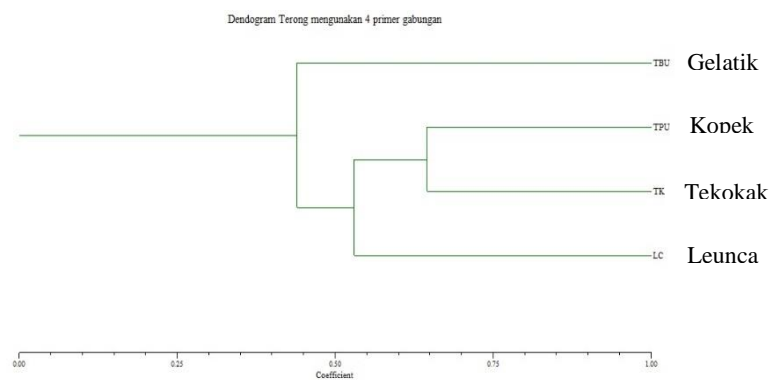

**Figure 2.** Dendrogram depicting the genomic relationship among four accessions of *Solanum melongena*.

Four *Solanum melongena* cultivars local from Indonesia namely leuca (LC), tekokak (TK), gelatik (TBU) and kopek (TPU) has been successfully justified with four primers. In early testing, Cetyl Trimethyl Ammonium Bromide (CTAB) was used to extract genomic DNA. CTAB is a common method of extraction of plant genomic DNA containing polysaccharides and polyphenolic compounds [18]. Basically, genomic DNA isolation consists of three major steps, namely the destruction of cell walls (lysis), separation of DNA from solid materials such as cellulose and proteins, and DNA purification [4]. In this study, our early stages we tested the quality and concentration of DNA. Quality testing and concentration on DNA isolation are important for the manipulation of all substances. DNA concentration and its purity can be determined by spectrophotometer using UV light with 260 and 280 nm [224]. The comparison in the test using UV light with a length of 260/280 nm gives a measure of the contamination of proteins in a DNA, while the use of UV light with a length of 260 nm aims to calculate the concentration of a sample DNA [17, 3]. UV absorption process could not be separate RNA contaminants, and contaminants such as proteins, phenols, oligo or polysaccharides contained in DNA are difficult to estimate [16, 4].

In general, the total amplification of the DNA genomes of the Solanaceae sample using selected primers yields a series of bands, among which there are bands commonly encountered throughout the sample and those of specific bands found only by specific species/cultivars. The overall band profile resulting from this RAPD primer amplification is the DNA of each type/cultivar. RAPD techniques are evaluated to distinguish individuals with close family ties [16, 13]. Based on phylogenetic tree Figure 2 showed that kopek (TPU) was closer to kinship with tekokak (TK).

The result of quantity test with spectrophotometer showed good result. Four accessions obtained DNA genomes with high purity and concentration. Primers used in this study were primers OPF-04, OPF-03, OPF-02, and OPF-01. Of the four primers used, all successfully amplify genomic DNA. At the time of testing the quality and quantity of DNA kopek (TPU) to get results that were not significant, electrophoresis results showed not too clear the DNA band cultivars. Measurements by using the speculator produce a pretty good number, there was DNA but the amount obtained was less. Inadequate quality and quantity test results do not guarantee that genomic isolation actually fails [9]. However, if there is too little DNA, then the test results may show negative results due to the minimum limitations of the tool [5]. Very few genomic DNAs so difficult to detect by means of equipment can still be used as templates in the PCR process [11]. PCR has the ability to multiply very powerful DNA fragments. PCR requires only a few templates to be able to reproduce certain DNA fragments [14, 24].

## 4. Conclusion
The dendrogram results showed that TPU (kopek eggplant) was closer to its hazard with TK (tekokak) with a similarity of 65%, whereas in LC (lenca) had 55% similarity, and in TBU (eggplant) had the furthest kinship with 49% equality.

**Acknowledgments**
This research was financially supported by the Research and Development Institutions (LEMLITBANG) Universitas Muhammadiyah Prof. Dr. HAMKA, Jakarta.